 \documentclass[twocolumn,aps,prl,showpacs,preprintnumbers,amsmath,amssymb]{revtex4}
\usepackage{epsfig} 

\begin{document}

\title{$^7$Be(p,$\gamma$)$^8$B astrophysical S-factor from precision cross section measurements}

\author{A. R. Junghans, E. C. Mohrmann,  K. A. Snover}
\author{T. D. Steiger}
\altaffiliation{present address: Cymer, Inc., 16750 Via del Campo Ct., San Diego, CA 92127.}
\author{E. G. Adelberger}
\author{ J. M. Casandjian}
\altaffiliation{permanent address: GANIL, B.P. 5027, 14021 Caen Cedex, France.}
\author{H. E. Swanson}
\affiliation{Center for Experimental Nuclear Physics and Astrophysics, University of  Washington,
Seattle,~Washington~98195}

\author{L. Buchmann}
\author{S. H. Park}
\altaffiliation{Dept. Physics, Seoul Nat. Univ., Seoul 151-742, Rep. of Korea.} 
\author{A. Zyuzin}

\affiliation{TRIUMF, 4004 Wesbrook Mall, Vancouver, B.C., Canada V6T 2A3}

\date{\today}

\begin{abstract}

We measured the $^7$Be(p,$\gamma$)$^8$B cross section from $\bar{E}_{cm}$ = 186 to 1200 keV, 
with a statistical-plus-systematic precision per point of better than $\pm 5\%$.  
All important systematic errors were measured including $^8$B backscattering losses.  
We obtain S$_{17}$(0) = 22.3 $\pm$ 0.7(expt) $\pm$ 0.5(theor) eV-b from our data at 
$\bar{E}_{cm} \leq$ 300 keV and the theory of Descouvemont and Baye.  
\end{abstract}
\pacs{26.20+f, 26.65+t, 25.40Lw}

\maketitle

It is now known that electron neutrinos ($\nu_e$'s) from the decay of $^8$B in the Sun 
oscillate into $\nu_{\mu}$'s and/or $\nu_{\tau}$'s, and possibly into sterile 
$\nu_x$'s~\cite{sno}.  The $\nu_e$ production rate is based on solar-model calculations 
that incorporate measured reaction rates for most of the solar burning steps, 
the most uncertain of which is the $^7$Be(p,$\gamma$)$^8$B rate.  
Improved production rate predictions are very important for limiting the allowed 
neutrino mixing parameters including possible contributions of sterile neutrinos.
The astrophysical S-factor S$_{17}$(0) for this reaction must be known to $\pm 5\%$   
in order that its uncertainty not be the dominant error in predictions of the 
solar $\nu_e$ flux~\cite{bahcall}. 
     
S$_{17}$(0) values based on previous direct measurements have quoted uncertainties of 
typically $\pm 9\%$ or 
larger~\cite{parker,kavanagh,vaughn,wiezorek,filippone,hammache,gialanella,strieder}
(see also the quoted $\pm5\%$ results of ref.~\cite{hass}), while for many of these 
experiments there are unsettled issues such as possible $^8$B backscattering losses.  
Indirect S$_{17}$(0) determinations based on Coulomb dissociation and 
peripheral transfer reactions are also available~\cite{davids}, 
but it is difficult to determine all of their important systematic errors.
  
We have made a precise determination of S$_{17}$(0) using a technique that incorporates 
several improvements over traditional methods.  We avoided a major difficulty in 
most previous experiments due to uncertain and nonuniform target areal density by using 
a $\sim$1 mm diameter beam magnetically rastered to produce a nearly uniform flux 
over a small $\sim$3 mm diameter target. We directly measured the energy loss profile 
of the target using a narrow $^7$Be($\alpha,\gamma)^{11}$C resonance and we determined
all important sources of systematic error including the first direct measurement 
of $^8$B backscattering losses.  

We used a 106 mCi $^7$Be metal target fabricated at TRIUMF and deposited on 
a molybdenum backing. The cross sections were measured using the 
University of Washington FN tandem accelerator with a terminal ion source.  
A proton beam, typically 10 $\mu$A, passed through an LN$_2$-filled 
cold trap directly upstream of the target.  Cryopumps were used for 
high-vacuum pumping, and sorption pumps for roughing.  The water-cooled 
target, and a plate with precision-sized circular apertures were mounted on 
opposite ends of a rotating arm. 
Rotating the arm 180$^{\circ}$ from its horizontal bombardment position 
placed a 3 mm aperture in the beam, and the target $\sim$4.5 mm from a 
450 mm$^2$ 40 micron Si  detector that counted $\beta$-delayed $\alpha$'s 
from $^8$B decay.  In each measurement, the arm was rotated through many complete cycles.  

We integrated 3 different beam currents: the current striking the target 
during the bombardment phase, and, during the $\alpha$-counting phase, 
the current striking the aperture and the current collected in a Faraday cup 
after passing through the aperture.  The target arm was biased to +300V.  
The neutral H content of the beam was found to be $<10^{-4}$, and the cup current 
changed by $< 0.5\%$ for a cup suppressor bias in the range --300$\pm$45V.
We estimated a $\pm 0.8\%$ beam flux integration uncertainty based on the 
difference of the good geometry (Faraday cup) and poor geometry (biased target arm) results.    
The beam was rapidly deflected from the target prior to and during arm movement.
The timing cycle intervals~\cite{filippone} were $t_1$ = $t_3$ = 1.50021 s, 
$t_2$= 0.24003 s and $t_4$=0.26004 s, and the (inverse) timing efficiency  
$\beta (^8$B)= 2.923 $\pm$ 0.005  assuming $t_{1/2}(^8$B) = 770 $\pm$ 3 ms~\cite{halflives}.

\begin{figure}
\unitlength1mm

\begin{picture}(80,128)
\put(-5,-2){\mbox{\epsfig{width=0.5\textwidth, angle=0, 
 file=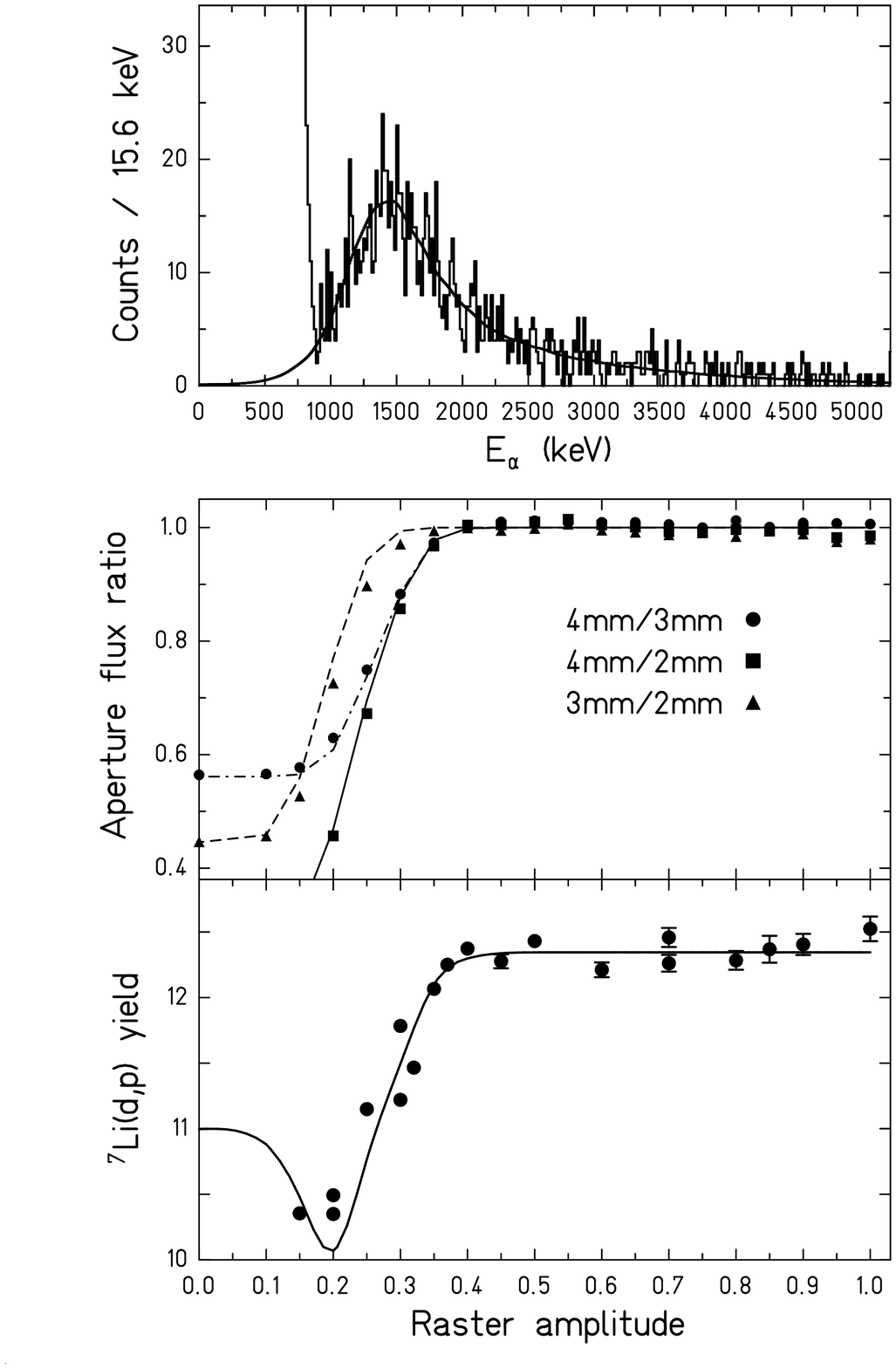}}}
\end{picture}

\caption{Top panel: $\alpha$-spectrum from $^7$Be(p,$\gamma$)$^8$B at 
$\bar{E}_{cm}$ = 186 keV. Middle panel: 770 keV deuteron beam transmission 
ratios through different apertures, vs. raster amplitude.  Bottom panel: 
$^7$Li(d,p)$^8$Li yield at 770 keV, normalized to the integrated beam flux through a 3 mm 
aperture, vs. raster amplitude, measured with the same tune as the aperture ratio data.      
 \label{dp}}
\end{figure}

In the limit of uniform beam flux, the $^7$Be areal density is unimportant
and the cross section is given by
\begin{equation}
\sigma(\bar{E}_{cm}) = \frac{Y_{\alpha}(E_p) F_{\alpha}(E_p) \beta(^8\mbox{B})}
{2N_p N_{\text{Be}}(t)
\Omega /4\pi}
\label{cross}
\end{equation} 
where $\bar{E}_{cm}$ is discussed below, $E_p$ is the bombarding energy, 
$Y_{\alpha}(E_p)$ is the $\alpha$ yield above a threshold energy of 895 keV, 
$F_{\alpha}(E_p)$ is a correction for the fraction of the $\alpha$-spectrum 
that lies below the threshold,  $N_p$ is the integrated number of protons per 
cm$^2$, $N_{\text{Be}}(t)$ is the number of $^7$Be atoms
and $\Omega$ is the solid angle of the $\alpha$-detector. 

In practice it is impossible to produce a completely uniform beam flux.  To understand 
the error associated with this approximation, one needs to know both the beam and 
target uniformities.  It is particularly important that the target be confined within 
a small central area.  This was insured by depositing the $^7$Be on a Mo backing 
consisting of a 4 mm diameter raised post surrounded by a mask tightly pressed 
around the post, with post plus mask machined flat as one piece.  After evaporation 
the mask was removed, eliminating unwanted tails on the $^7$Be radial 
distribution~\cite{targetpaper}.  

\begin{figure}
\unitlength1mm
\begin{picture}(80,113)
\put(-5,-2){\mbox{\epsfig{width=0.5\textwidth, angle=0, 
 file=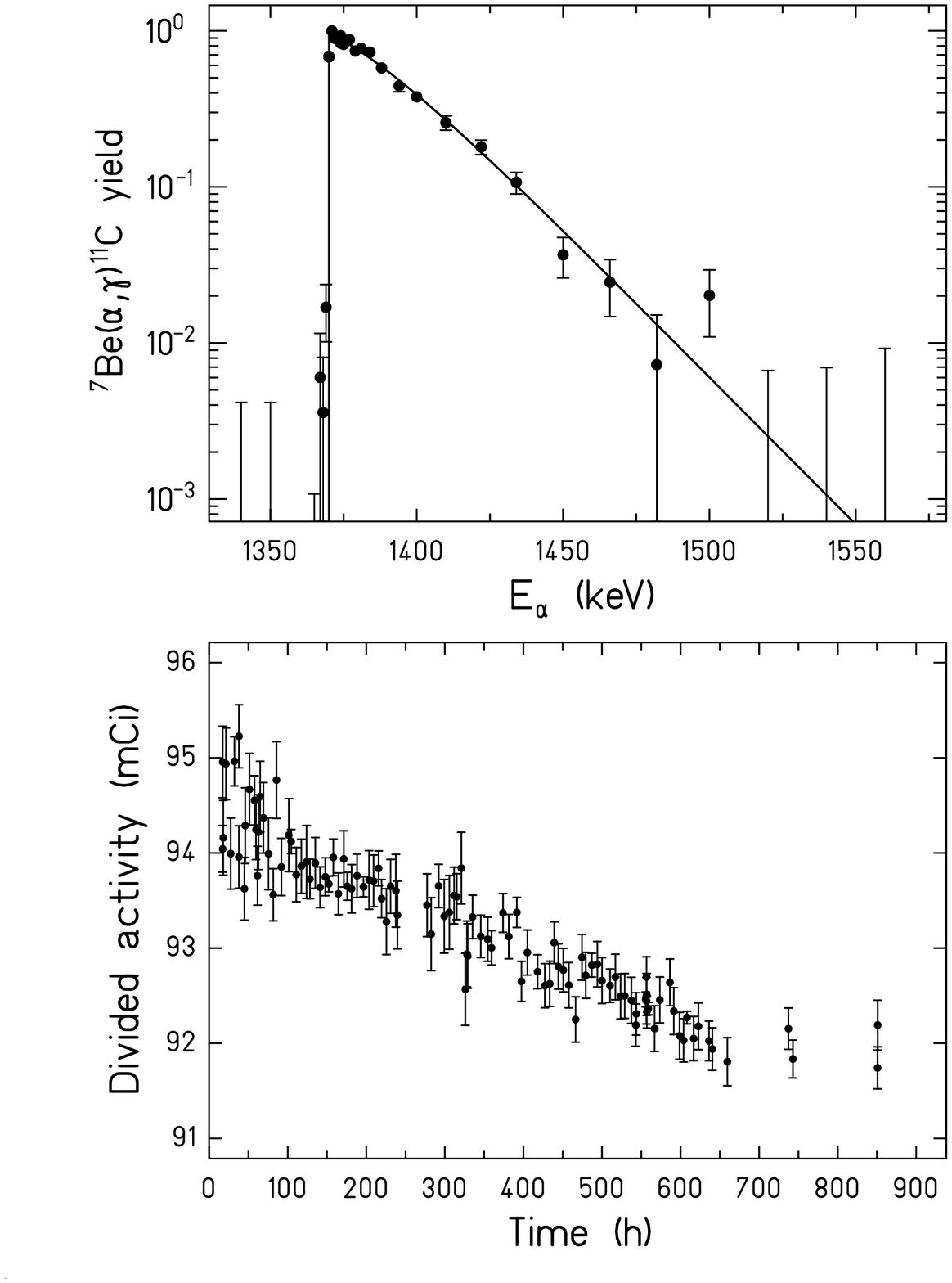}}}
\end{picture}
    \caption{Top panel: $^7$Be($\alpha, \gamma)^{11}$C resonance profile corrected 
for small backgrounds due to cosmic rays and a nonresonant yield from $^9$Be($\alpha,n)^{11}$C.  
Bottom panel: $^7$Be activity divided by the $^7$Be decay curve, showing sputtering losses.
    \label{activity}}
\end{figure}

The beam uniformity was determined by measuring the transmissions through 
2, 3 and 4 mm apertures as functions of the (equal) amplitudes of the x and y 
triangular raster waveforms.  Fig.~\ref{dp} shows measurements
with a 770 keV deuteron beam, and curves calculated by folding a Gaussian with 
a rectangular function.  
The uniformity of the product of the beam and target densities was determined by 
the raster-amplitude dependence of the $^7$Li(d,p)$^8$Li yield from the $^7$Be 
target at $E_d$= 770 keV, shown in Fig.~\ref{dp}. The curve is a 1-parameter 
folding of the target density estimated from $\gamma$-activity scans, and beam 
profile determined by the transmission ratios, including a fitted target-aperture
misalignment of 0.5 mm. The point at which this yield flattened out determined 
the minimum safe raster amplitude, and is similar to the point at which the 
aperture ratio data flattened out.  We chose 0.42 as the safe raster amplitude for 770 keV
deuterons, and assigned a conservative $\pm 1\%$ nonuniformity uncertainty here.  
Aperture transmission curves, measured at most proton energies, determined the 
minimum raster amplitude for each energy and tune for which the beam-target 
nonuniformity was $< 1\%$. Independent estimates of the safe raster amplitudes
were made by folding the target density distribution~\cite{targetpaper} 
with beam-flux distributions determined from the proton aperture-transmission data.  

$N_{\text{Be}}(t)$  was determined with the target arm vertical by counting 478 keV 
$\gamma$-rays  $\it{in}$ $\it{situ}$ using a collimated Ge detector located on 
top of the target chamber.  We assumed $t_{1/2}$ = 53.12 $\pm$ 0.07d~\cite{halflives} 
and a 10.52 $\pm 0.06\%$ branch~\cite{halflives} to the 478 keV level.  
The Ge efficiency $\epsilon_{478}$ was determined to $\pm 1.3\%$ from a fit
to 14 lines from $^{125}$Sb, $^{134}$Cs, $^{133}$Ba, $^{137}$Cs and $^{54}$Mn 
sources calibrated typically to $\pm 0.8\% (1\sigma)$~\cite{isotope}, 
with $\chi^2 / \nu$ = 2.2. We obtained a second $^{137}$Cs source calibrated 
independently to $\pm 0.4\% (1\sigma)$~\cite{french}. 
The relative activity of the two $^{137}$Cs sources agreed within $\pm 0.1\%$.   
As can be seen in Fig.~\ref{activity}, 2.5 mCi of $^7$Be was lost due to beam 
sputtering during the cross section measurements.  

\begin{table}[h]
\caption{Percent uncertainties $\Delta$S$_{17}$/S$_{17}$.}\label{table1}
\begin{tabular}{l r}
\hline \hline
Statistical errors & 1.0-2.8 \\
    \hline
Varying systematic errors: \\
\hspace{0.5cm}	proton energy calibration & 0.2-0.6 \\
\hspace{0.5cm}	target thickness & 0.0-1.0 \\
\hspace{0.5cm}	target composition & 0.0-1.1 \\
   \hline 
Scale factor errors: \\
\hspace{0.5cm}	beam-target inhomogeneity & 1.0\\
\hspace{0.5cm}	integrated beam flux & 0.8\\
\hspace{0.5cm}	target activity & 1.9\\
\hspace{0.5cm}	solid angle & 1.2\\
\hspace{0.5cm}	$\alpha$-spectrum cutoff & 0.7\\
\hspace{0.5cm}	backscattering & 0.5\\
\hspace{0.5cm}	timing cycle & 0.2\\
Total scale factor error & 2.7\\
\hline \hline
\end{tabular}
\end{table}

We inferred $\Omega$ with the aid of a ``far" Si detector~\cite{filippone} 
located 47.42 $\pm$ 0.09 mm from the target and collimated to an area of 
248.8 $\pm$ 0.4 mm$^2$.  From geometry, $\Omega_{far}$ = 0.1078 $\pm$ 0.0004 sr, 
where the zero of the distance scale was checked using a $^{148}$Gd $\alpha$-source.  
$\Omega/\Omega_{far}$ was determined using the $^7$Li(d,p)$^8$Li reaction.  
A differential correction for $\alpha$-particles lost below the threshold was applied 
based on the (d,p) angular distribution~\cite{dpangdist} and SRIM~\cite{srim} calculations 
including $^8$Li-straggling. We obtained $\Omega$ = 3.82 $\pm$ 0.04 sr.  
This result was checked using different detectors and different size collimators 
for $\Omega_{far}$. 

The yields $Y_{\alpha}(E_p)$ were corrected for a small beam-off background (3.9\% 
at the lowest E$_p$). The beam-related background was checked at several energies 
and found to be negligible. The $\alpha$-spectrum cutoff factors for 
$^7$Be(p,$\gamma)^8$B were estimated from SRIM calculations, 
including straggling, fitted to 23 different spectra.   $F_{\alpha}(E_p)$ varied 
linearly from $1.039 \pm 0.007 $ at E$_p$ = 221 keV to $1.086 \pm 0.008 $ at 1379 keV.
The accelerator energy calibration was determined to $\pm 0.17\%$ from 
$^{19}$F(p,$\alpha \gamma)^{16}$O resonances at $E_p$ = 340.46 $\pm$ 0.04, 
483.91 $\pm$0.10 and 872.11 $\pm$ 0.20 keV~\cite{f19pag}.

Corrections for energy averaging of the proton beam due to finite target thickness 
are important, particularly at low $E_p$.  We directly measured the
beam energy loss profile in the target using the narrow ($\Gamma << $ 1 keV)
$^7$Be($\alpha, \gamma)^{11}$C resonance~\cite{be7ag} which we found 
at E$_{\alpha}$ = 1378 $\pm$ 3 keV. The mean $\alpha$-energy loss was 26 $\pm$ 2 keV, 
based on the average of three measurements, one of which is shown in Fig.~\ref{activity}.
The excellent reproducibility of the apparent $^7$Be($\alpha,\gamma$)$^{11}$C
resonance energy measured in the middle of, and after the $^7$Be(p,$\gamma$)$^8$B
measurements ($\Delta E_{\alpha}$ = 1 $\pm$ 3 keV), indicated negligible 
carbon buildup and target damage due to bombardment.

An important error in some previous experiments was loss of $^8$B from the 
target due to backscattering (and loss of $^8$Li when $^7$Li(d,p)$^8$Li 
was used for absolute cross section normalization)~\cite{weissman,strieder2}.  
These losses may be sizeable when a high-Z backing is used, or if there are 
high-Z contaminants in the target.  We made the first direct measurements of
$^8$B backscattering losses in the $^7$Be(p,$\gamma$)$^8$B reaction using 
our $^7$Be target in a fixed mount, and large-diameter water-cooled Cu 
catcher plates on each end of the rotating arm. A 4 mm hole in the center 
of each plate allowed the beam to pass through.  We found small backscattering
losses of 1.3 $\pm$ 0.3$\%$ and 0.9 $\pm$ 0.2$\%$  at $E_p$ = 724 and 1379 keV, 
respectively, and made a constant 1.0 $\pm$ 0.5$\%$ correction to 
our data for this effect.
\begin{figure}[h]
\unitlength1mm
\begin{picture}(80,75)
\put(-10,-5){\mbox{\epsfig{width=0.5\textwidth, angle=0, 
file=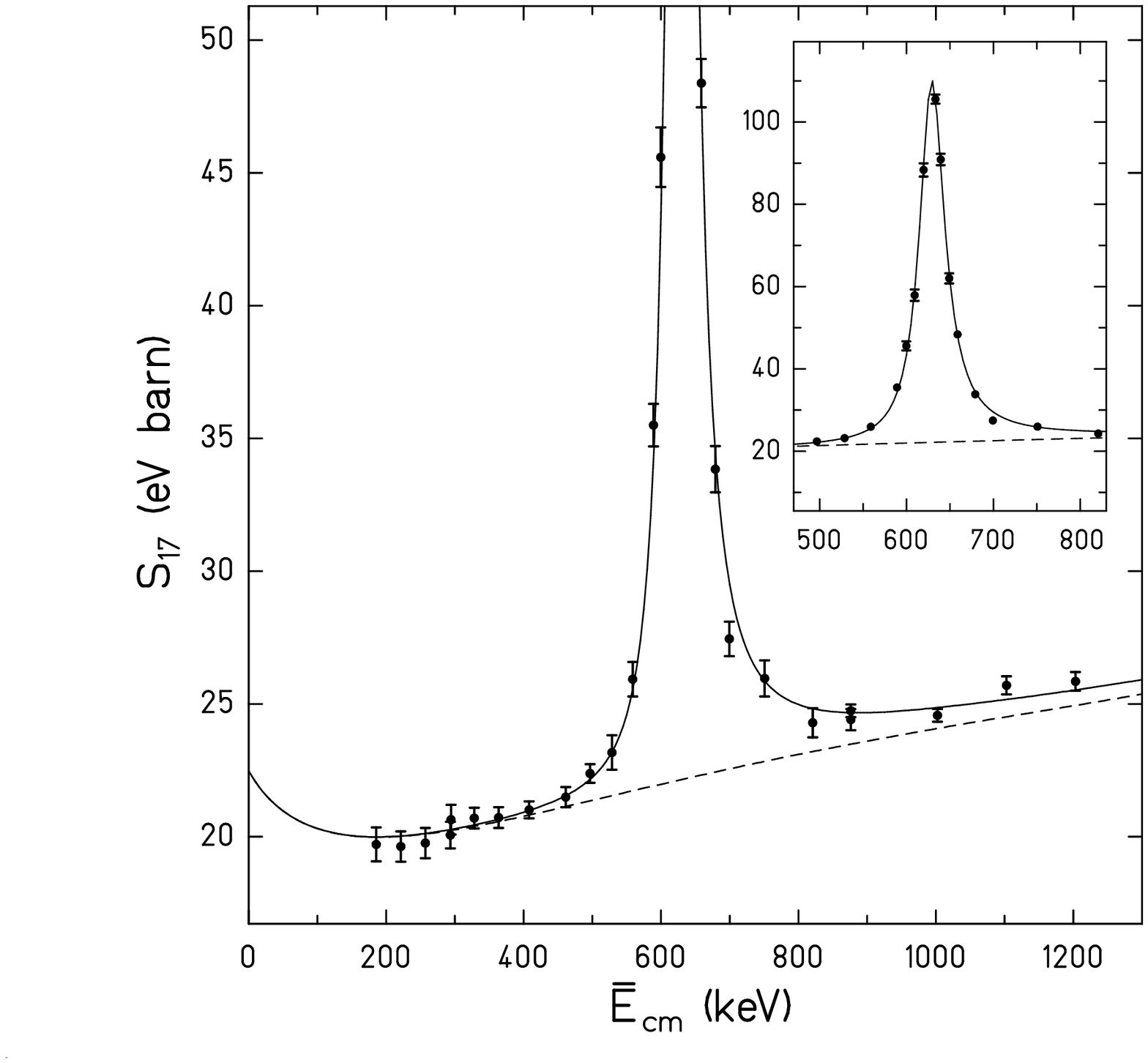}}}
\end{picture}
\caption{S$_{17}(\bar{E}_{cm}$) vs. $\bar{E}_{cm}$ from this work.  
Error bars are statistical plus $\it{varying}$ systematic errors.  
Solid curve: DB theory plus a Breit-Wigner resonance.  Dashed curve: 
DB theory.  Inset: resonance region.
\label{ourSfactor}}
\end{figure}

Fig.~\ref{ourSfactor} shows our S-factors calculated from the relation 
S$_{17}(\bar{E}_{cm}) = \sigma(\bar{E}_{cm})\bar{E}_{cm}
\mbox{exp}[(E_G/\bar{E}_{cm})^{1/2}]$ 
(see e.g.~\cite{filippone}) with $E_G$ = 13799.3 keV.  We computed $\bar{E}_{cm}$ 
by inverting the expression $\sigma(\bar{E}_{cm}) = \bar{\sigma}$, 
where $\bar{\sigma}$ was obtained by fitting the cross section data 
including averaging over the target profile, and $\sigma$ is the corresponding
unaveraged cross section. These $\bar{E}_{cm}$ values are very close to the mean
proton energy in the target, except near the resonance where they differ by $<1\%$.
Fig.~\ref{ourSfactor} also shows a fit to all our data of the (scaled) cluster model 
theory of Descouvemont and Baye~\cite{db} plus an $\bar{E}_{cm}$ = 
630 $\pm$ 2 keV Breit-Wigner resonance (with energy-dependent 
$\Gamma_p$ and $\Gamma_{\gamma}$). This fit yields S$_{17}$(0) = 
22.5 $\pm$ 0.6 eV-b and $\chi^2 / \nu$ = 1.3 ($\nu$=25)~\cite{newthresh}, 
where the quoted uncertainty includes the scale factor error 
of $\pm$ 2.7 $\%$ (Table~\ref{table1}). Fits with other theories~\cite{jennings} 
did not reproduce our measured energy dependence as well ($\chi^2 / \nu$ = 1.7-16).

The theoretical uncertainty in the energy dependence of S$_{17}$ decreases 
with beam energy below the resonance, as the capture becomes increasingly 
extranuclear.  Therefore it is important to determine S$_{17}$(0) 
from low energy data.
Fitting the DB theory to our data at $\bar{E}_{cm} \leq$ 300 keV we find 
S$_{17}$(0) = 22.3 $\pm$ 0.7 eV-b and $\chi^2 / \nu$ = 0.3 (here, as above, 
the error includes statistical plus systematic contributions).
In addition, there is an extrapolation uncertainty, which has been estimated 
to be as small as $\pm$ 0.2 eV-b~\cite{jennings}, and which we estimate 
conservatively as $\pm$ 0.5 eV-b from the rms deviation of 11 different 
theoretical fits to our data for $\bar{E}_{cm} \leq$ 300 keV~\cite{alltheories}.
Thus our final result is
\begin{equation}
\mbox{S}_{17}(0) = 22.3 \pm 0.7\mbox{(expt)} \pm 0.5\mbox{(theor)} 
\hspace{0.2cm} \mbox{eV-b.}
\label{S17}
\end{equation} 
\begin{figure}[h]
\unitlength1mm
\begin{picture}(80,50)
\put(-10,-10){\mbox{\epsfig{width=0.5\textwidth, angle=0, 
file=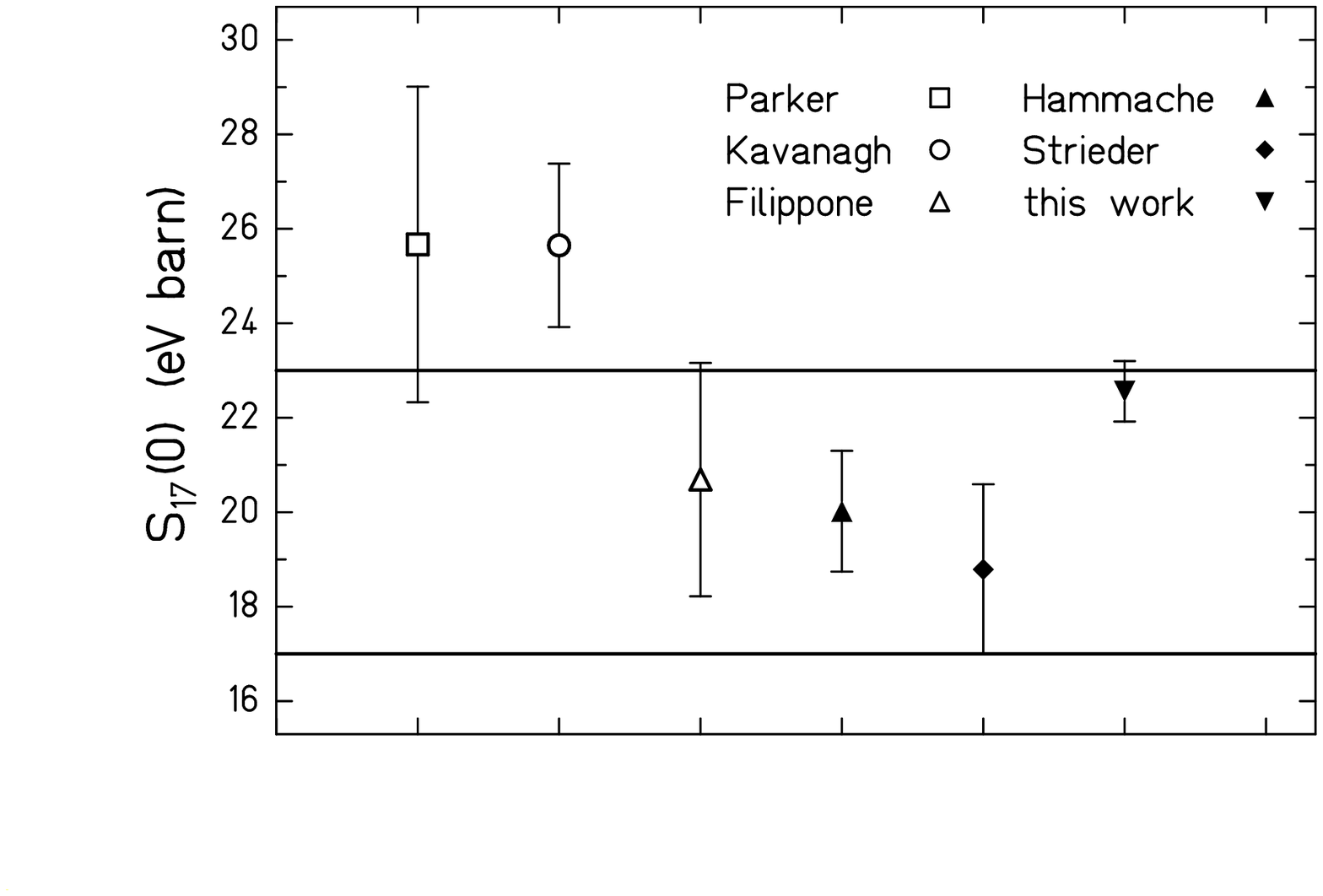}}}
\end{picture}
\caption{S$_{17}$(0) from our fits of the DB theory to $\bar{E}_{cm} \leq$ 425 keV data 
from this and previous measurements. Horizontal lines indicate the 
19 $^{+ 4}\!\!\!\!\!\!_{- 2}$ eV-b range recommended by~\protect\cite{adelberger}. 
Fits over a wider $E_p$-range give similar results but with smaller errors for 
other experiments. 
\label{allSfactors}}
\end{figure} 

In order to compare all direct measurements below the resonance, we made DB fits to all 
data at $\bar{E}_{cm} \leq$ 425 keV -- this work 
and~\cite{parker,kavanagh,filippone,hammache,strieder,nndc} renormalized 
to $\sigma$[$^7$Li(d,p)$^8$Li] = 152 $\pm$ 6 mb~\cite{dpaverage} where appropriate. 
The results are shown in Fig.~\ref{allSfactors}.  
Results from~\cite{parker,kavanagh,filippone} may suffer additional error 
from $^8$B and $^8$Li backscattering losses; in \cite{hammache}, calculated 
corrections were applied, while in \cite{strieder}, a low-Z backing was used 
and losses were assumed negligible. 

We have reduced the error on S$_{17}$(0) so that it no longer dominates the 
uncertainty in the calculated solar $^8$B $\nu_e$ production rate.
While our S$_{17}$(0) value agrees within errors with the previously recommended 
value of 19 $^{+ 4}\!\!\!\!\!\!_{- 2}$ eV-b~\cite{adelberger}, it is 17\% larger.  
Thus 17\% more of the $^8$B solar $\nu_e$'s oscillate into other species than 
given in ref.~\cite{bahcall}. 

We thank the staff of CENPA and TRIUMF, N. Bateman, R. O'Neill, J. Martin 
and R. Hoffenberg for their help, J.N. Bahcall for helpful comments, 
and the U.S.D.O.E., Grant $\#$DE-FG03-97ER41020, and the N.S.E.R.C. of Canada 
for financial support.


\begin{thebibliography}{}
\bibitem{sno}  Q.R. Ahmad et al., Phys. Rev. Lett. {\bf 87}, 071301-1, 2001.
\bibitem{bahcall} J. N. Bahcall, S. Basu and M.H. Pinsonneault, Phys. Lett. B {\bf 433},
 1, 1998; Astrophys. J. {\bf 555}, 990, 2001.
\bibitem{parker} P.D. Parker, Phys. Rev. {\bf 150}, 851, 1966, and private comm.    2001.
\bibitem{kavanagh} R.W. Kavanagh, Bull. Am. Phys. Soc. {\bf 14}, 1209 (1969) and 
                    private comm., 2001.
\bibitem{vaughn} F.J. Vaughn et al., Phys. Rev. C {\bf 2}, 1657, 1970.
\bibitem{wiezorek} C. Wiezorek et al., Zeits. f\"u{}r Physik A {\bf 282}, 121, 1977.
\bibitem{filippone} B.W. Filippone et al., Phys. Rev. C {\bf 28}, 2222, 1983.
\bibitem{hammache} F. Hammache et al., Phys. Rev. Lett. {\bf 80}, 928, 1998; 
F. Hammache, Ph.D. thesis, 1999; F. Hammache et al., Phys. Rev. Lett. {\bf 86}, 3985, 2001.
Our DB fits to the low and high energy data sets of the PRL2001 paper 
yield S$_{17}$(0) values 6\% and 4\% higher than quoted there.
\bibitem{gialanella} L. Gialanella et al., Eur. Phys. J. A {\bf 7}, 303, 2000.
\bibitem{strieder} F. Strieder et al., Nucl. Phys. A., in press, 2001.
\bibitem{hass} M. Hass et al., Phys. Lett. B {\bf 462}, 237, 1999.
\bibitem{davids} B. Davids et al., Phys. Rev. Lett. {\bf 86}, 2750, 2001; 
A. Azhari et al., Phys. Rev. C {\bf 63}, 055803/1, 2001, and refs. therein.
\bibitem{halflives} S.Y.F. ~Chu, ~L.P. ~Ekstrom and R.B. Firestone: 
nucleardata.nuclear.lu.se/NuclearData/toi/
\bibitem{targetpaper} A. Zyuzin et al., Nucl. Inst. Meth. B, in press.
\bibitem{isotope} Isotope Products Corp., Burbank CA, USA.
\bibitem{french} CERCA and LNHB, CEA, Saclay, France.
\bibitem{dpangdist} A.J. Elwyn et al., Phys. Rev. C {\bf 25}, 2168, 1982.
\bibitem{srim} J.F. Ziegler, www.srim.org.
\bibitem{f19pag} D.R. Tilley et al., Nucl. Phys. A {\bf 636}, 249, 1998.
\bibitem{be7ag} G. Hardie et al., Phys. Rev. C {\bf 29}, 1199, 1984.
\bibitem{weissman} L. Weissman et al., Nucl. Phys. A {\bf 630}, 678, 1998.
\bibitem{strieder2} F. Strieder et al., Eur. Phys. J. A {\bf 3}, 1, 1998.
\bibitem{db} P. Descouvemont and D. Baye, V2 interaction, Nucl. Phys. A {\bf 567}, 341, 1994, 
and private comm. 2001. 
\bibitem{newthresh} A separate analysis based on an alpha threshold energy of 1000 keV 
agrees with this result within 0.7\%. 
\bibitem{jennings} B.K. Jennings, S. Karataglidis and T.D. Shoppa, 
Phys. Rev. C {\bf 58}, 3711, 1998.
\bibitem{alltheories} DB, B1, B2, C2B, C8B, 1.0 and 2.4 as defined in 
\protect\cite{jennings}, plus: C. Johnson et al., ApJ {\bf 392}, 320, 1992; 
F. Nunes et al., Nucl. Phys. A {\bf 634}, 527, 1998; 
K. Bennaceur et al. Nucl. Phys. A {\bf 651}, 289, 1999; and private comms.
\bibitem{nndc} EXFOR data base: www.nndc.bnl.gov/nndc/exfor.
\bibitem{adelberger} E.G. Adelberger et al., Rev. Mod. Phys. {\bf 70}, 1265, 1998.
\bibitem{dpaverage} Weighted average of values given in~\protect\cite{weissman,adelberger}.
\end{thebibliography}
\end{document}